\documentclass[11pt]{article}
\usepackage{moriond,epsfig}

\def\Journal#1#2#3#4{{#1} {\bf #2}, #3 (#4)}

\def\NIMA{{\em Nucl. Instrum. Methods} A}

\def\PLB{{\em Phys. Lett.}  B}
\def\PRL{\em Phys. Rev. Lett.}
\def\PRD{{\em Phys. Rev.} D}

\def\be{\begin{equation}}
\def\ee{\end{equation}}
\def\bea{\begin{eqnarray}}
\def\eea{\end{eqnarray}}

\begin{document}
\vspace*{4cm}
\title{Search for new physics in {\it B} to {\it VV} decays and other hot topics from Belle}

\author{ Katsumi Senyo, for Belle Collaboration }

\address{Department of Physics, Nagoya University,\\
  Furo-cho, Chikusa-ku, Nagoya, Japan}

\maketitle\abstracts{
We report studies in polarization in $B$ decay into two vector mesons
with data equivalent to $253 {\rm fb^{-1}}$ on $\Upsilon(4S)$ resonance at KEKB.
In $B^0 \to \phi K^{*0}$, $B^+ \to \phi K^{*+}$ and
$B^+ \to \rho^+ K^{*0}$ decays, we determine
$f_L = 0.45 \pm 0.05({\rm stat.}) \pm 0.02({\rm syst.})$,
$f_L = 0.52 \pm 0.08({\rm stat.}) \pm 0.03({\rm syst.})$
and $f_L = 0.43 \pm 0.11({\rm stat.}) ^{+ 0.05}_{-0.02}({\rm syst.})$ respectively,
where $f_L$ is a ratio between longitudinal and transverse polarization,
and a na\"{i}ve theoretical estimation assumes $f_L \sim 1$.
The discrepancy from 1 in $f_L$ may suggest existence of new amplitude
within or/and beyond the Standard Model.
}

\section{Introduction}
Na\"{i}ve factorization in the Standard Model (SM) predicts that the longitudinal
polarization fraction ($f_L$) in $B$ meson decays to light vector-vector($VV$)
final states is close to unity~\cite{kagan}.  In the tree dominated
$B^+ \to \rho^+\rho^0$ and $B^0 \to \rho^+\rho^-$ decays, this prediction
has been confirmed~\cite{zhang,babarrhorho1,babarrhorho2,babarphikstar}.  In the contrast,
for the pure $b \to s$ penguin $B \to \phi K^*$ decay, Belle\cite{chen} and
{\sc BaBar}\cite{babarphikstar} have found that longitudinal and transverse
polarization fraction are comparable, which is in disagreement with the
factorization expectation.  Possible explanations for this discrepancy include
enhanced non-factorizable contributions such as penguin annihilation\cite{kagan},
large $SU(3)$ breaking in form factors\cite{hnli}, or new physics\cite{grossman,yang}.
It is therefore important to perform polarization measurements 
in $B \to \phi K^*$ with larger data set and in other
$VV$ modes, in particular, in the pure penguin $b\to s \bar{d}d$ decay
$B^+ \to \rho^+ K^{*0}$.

In this study, we use a amount of $253 {\rm fb^{-1}}$ of data on $\Upsilon(4S)$
resonance, equivalent to approximately $274.5\times 10^6 B\bar{B}$ pairs,
collected by Belle detector\cite{belle} at KEKB $e^+e^-$ collider\cite{kekb}.
Detailed description of Belle detector is found in elsewhere\cite{belle}.

\section{$B \to \phi K^*$ Analysis}\label{btophik}
The event reconstruction is performed as follows; Candidate $B$ mesons
are reconstructed from $\phi$ and $K^*$ candidates and are identified
by the energy difference $\Delta E = E^{\rm cms}_B - E^{\rm cms}_{\rm beam}$,
the beam constrained mass $M_{\rm bc} = \sqrt{(E^{\rm cms}_{\rm beam})^2 - (p^{\rm cms}_B)^2}$, and $K^+K^-$ invariant mass ($M_{K^+K^-}$).
$E^{\rm cms}_{\rm beam}$ is the beam energy in the center-of-mass~(cms), and
$E^{\rm cms}_B$ and $p^{\rm cms}_B$ are the cms energy and momentum of
the reconstructed $B$ candidate.  The $B$-meson signal region is defined
as $M_{\rm bc} > 5.27 {\rm GeV}/c^2$, $|\Delta E| < 45 {\rm MeV}$, and
$|M_{K^+K^-} - M_\phi| < 10 {\rm MeV}/c^2$.  The invariant mass of the $K^*$
candidate is required to be less than $70 {\rm MeV}/c^2$ from the nominal $K^*$
mass.  The signal region is enlarged to $-100{\rm MeV} , \Delta E < 80
{\rm MeV}$ for $B^+ \to \phi K^{*+}(K^{*+} \to K^+\pi^0)$ because of the impact
of shower leakage on $\Delta E$ resolution.  The additional requirement
$\cos \theta_{K^*} < 0.8$ is applied to reduce low momentum $\pi^0$ background,
where $\theta_{K^*}$ is the angle between the direction opposite to the $B$ and the daughter kaon in the rest frame of $K^*$.  

The dominant background is $e^+e^- \to q\bar{q} (q = u, d, c, s)$ continuum
production.  Several variables including $S_{\perp}$\cite{sperp}, the thrust
angle, and the modified Fox-Wolfram moments~\cite{sfw} are used to exploit
the differences between the event shapes for continuum $q\bar{q}$ production~(jet-like) and for $B$ decay~(spherical) in the cms frame of the $\Upsilon(4S)$.
These variables are combined into a single likelihood ratio $\mathcal{R}_s
= \mathcal{L}_s / (\mathcal{L}_s + \mathcal{L}_{q\bar{q}}$, where
$\mathcal{L}_s$($\mathcal{L}_{q\bar{q}}$) denotes the signal ( continuum) likelihood.
The selection requirements on $\mathcal{R}_s$ are determined by maximizing
the value of $N_s / \sqrt{N_s + N_b}$ in each $B$-flavor-tagging quality
region\cite{ftag}, where $N_s$($N_b$) represents the expected number of signal
(background) events in the signal region.

Backgrounds from other $B$ decay modes such as $B \to K^+K^-K^*$,
$B \to f_0(980)K^*(f_0 \to K^+K^-)$, $B\to \phi K \pi$, $B \to K^+K-K\pi$,
and cross-feed between the $\phi K^*$ and $\phi K$ decay channels are studied
by Monte Carlo(MC) simulations.
The uncertainty in the $f_0(980)$ width (40-100 ${\rm MeV}/c^2$)\cite{f02} is
taken as a source of systematic error.  The contribution from $B \to K^+K^-K^*$
($B \to f_0 K^*$) is estimated to be 1-7\%(1-3\%) of the signal yield.
The background from $B\to \phi K \pi$ decays is evaluated with fits
to the $K\pi$ invariant mass and is found to be about 1\%.  To remove
the contamination from $\phi K$ decays, these decays are explicitly
reconstructed and rejected.

The signal yields ($N_S$) are extracted by extended unbinned maximum-likelihood
fits performed simultaneously to the $\Delta E$, $M_{bc}$ and $M_{K^+K^-}$
distributions.  Reconstructed $B$ candidates with $|\Delta E| < 0.25{\rm GeV}$,
$M_{bc}>5.2 {\rm GeV}/c^2$, and $M_{K^+K^-} <1.07 {\rm GeV}/c^2$ are included
in the fits.  The signal probability density functions (PDFs) are products of
Gaussians in $\Delta E$ and $M_{bc}$, and a Breit-Wigner shape in $M_{K^+K^-}$.
Bifurcated Gaussians(Gaussians with different widths on either side of the
mean) are added to model the tails in the $\Delta E$ distributions.
The means and widths of $\Delta E$ and $M_{bc}$ are verified using
$B \to J/\psi K^*$ decays.  The mean and width of the $\phi$ mass peak
are determined using an inclusive $\phi \to K^+K^-$ data sample.

\begin{figure}
  \begin{minipage}{0.45\hsize}
  \begin{center}
\psfig{figure=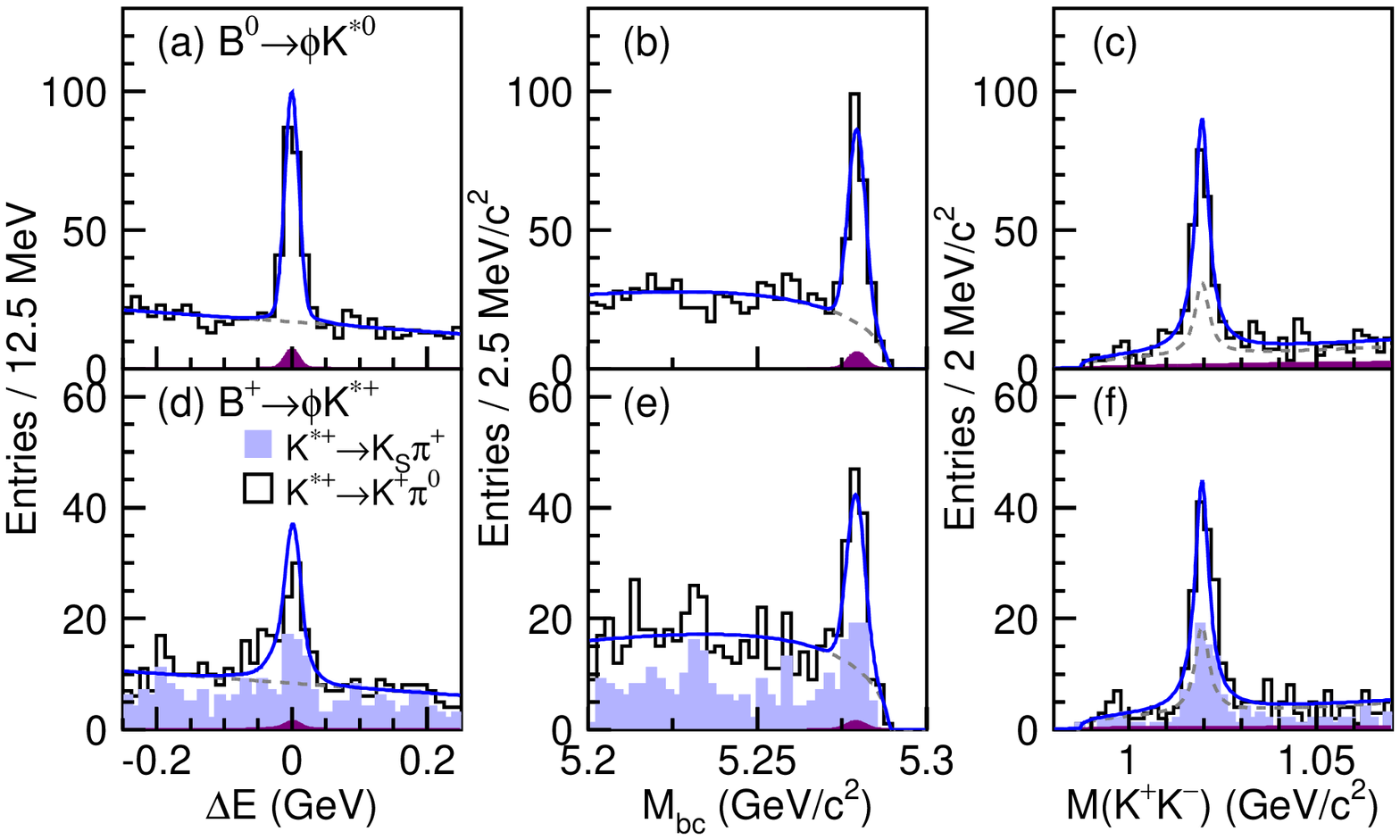,height=1.6in}
  \end{center}
\caption{Distributions of the $\Delta E$, $M_{bc}$ and $M_{K^+K^-}$ for $B^0
\to \phi K^{*0}$((a), (b) and (c)), and for $B^+ \to \phi K^{*+}$((d), (e)
and (f)), with other variables in signal region.  Solid curves show the
fit results.  The continuum background components are shown
by the dashed curves.  The dark shaded areas represent the contributions
from $B \to K^+K^-K^*$ and $B \to f_0 K^*$ decays.
\label{fig:phikstar1}}
\end{minipage}
\begin{minipage}{0.45\hsize}
 \begin{center}
\psfig{figure=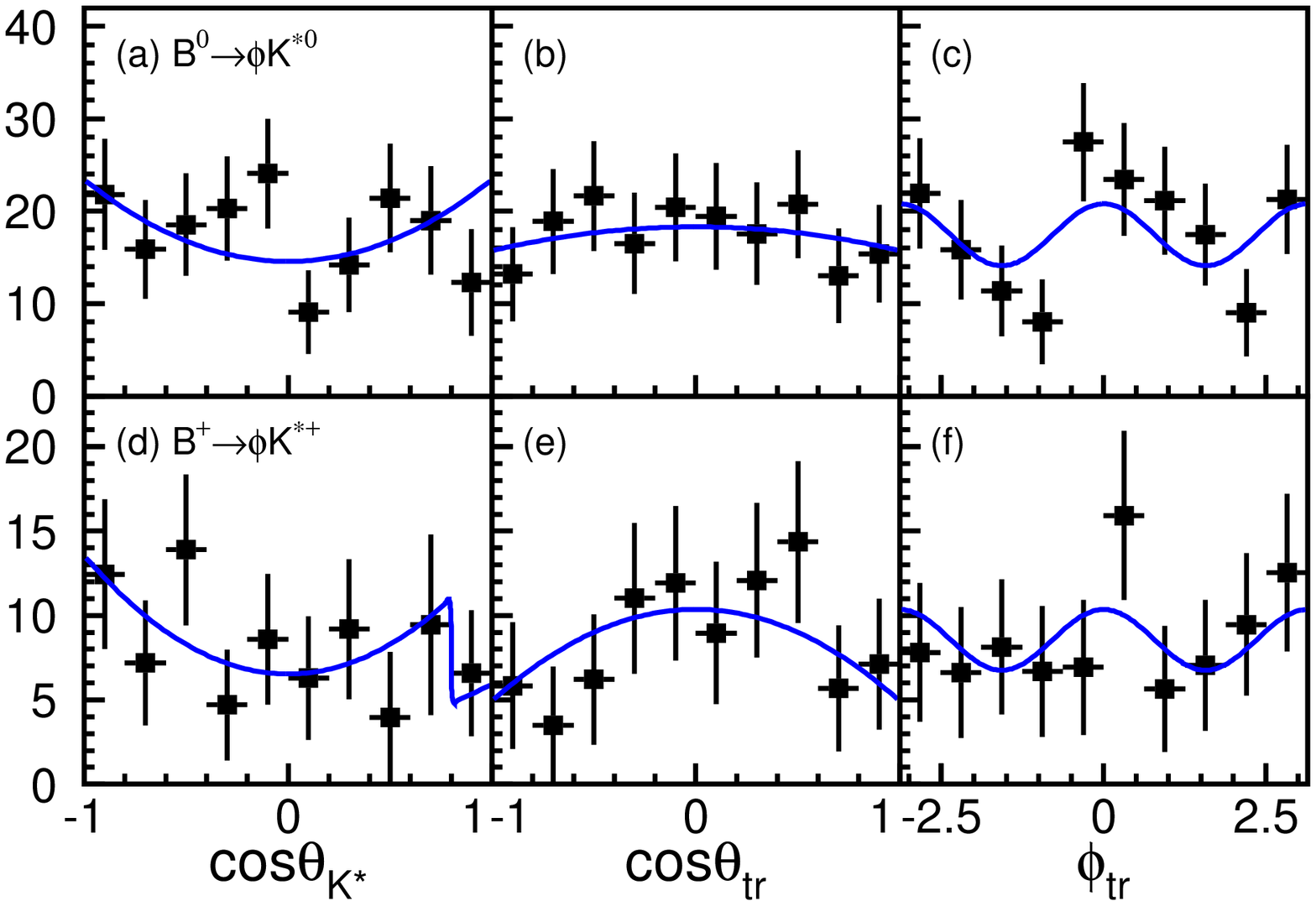,height=1.6in}
\end{center}
\caption{Projected distributions of the three transversity angles for
$B^0\to \phi K^{*0}$((a), (b) and (c)), and for $B^+\to \phi K^{*+}$
((d), (e) and (f)).  Solid lines show the fit results.  The points with
error bars show the efficiency corrected data after background subtraction.
The two $K^{*0}$ decay modes are combined in (d), (e) and (f).
The discontinuity in (d) is due to the requirement of $\cos\theta_{K^*}$
in $B^+\to \phi K^{*+}(K^{*+} \to K^+\pi^0)$.\label{fig:angdist}}
 \end{minipage}
\end{figure}

The PDF shapes for the continuum events are parameterized by and ARGUS
function\cite{argus} in $M_{bc}$, a linear function in $\Delta E$, and a sum
of a threshold function and a Breit-Wigner function in $M_{K^+K^-}$.
The parameters of the functions are determined by a fit to the events
in the sideband.  The signal and background yields are allowed to float
in the fit while other PDF parameters are fixed.  The direct $CP$ asymmetries,
$A_{CP} = \frac{N(\bar{B}\to \bar{f})-N(B\to f)}{N(\bar{B}\to \bar{f})+N(B\to f)}$,
 are also studied.  The measured signal yields and direct $CP$ asymmetries are summarized in Table~\ref{phikyield}.  The distributions of $\Delta E$, $M_{bc}$ and $M_{K^+K^-}$ are shown in Figure~\ref{fig:phikstar1}.

\begin{table}
\caption{Number of events observed in the signal region($N_{ev}$) , signal yields
($N_s$), and the direct $CP$ asymmetries($A_{CP}$) obtained in the fits, with
  statistical and systematic uncertainties.\label{phikyield}}
\begin{center}
\begin{tabular}{l|ccc}
\hline\hline
Mode & $N_{ev}$ & $N_s$ & $A_{CP}$\\
\hline
$\phi K^{*0}$ & $309$ & $173 \pm 16$ & $ 0.02 \pm 0.09 \pm 0.02 $  \\
\hline
$\phi K^{*+}$ & $173$ & $85^{+16}_{-11}$ & $ -0.02 \pm 0.14 \pm 0.03 $  \\
$\phi K^{*+}(K^0_S\pi^+)$ & $76$ & $37.9^{+7.7}_{-7.0}$ & $ -0.14 \pm 0.21 \pm 0.04 $  \\
$\phi K^{*+}(K^+\pi^0)$ & $97$ & $47.3^{+9.1}_{-8.1}$ & $ 0.09 \pm 0.19 \pm 0.04 $  \\
\hline\hline
\end{tabular}
\end{center}
\end{table}
\begin{table}
\caption{The decay amplitudes obtained for $B^0\to \phi K^{*0}$ and
$B^+\to\phi K^{*+}$.  the first uncertainties are statistical and the
second are systematic. \label{angresults}}
\begin{center}
\begin{tabular}{c|cc}
\hline\hline
 $N_{ev}$ & $\phi K^{*0}$ & $\phi K^{*+}$\\
\hline
$|A_0|^2$ & $0.45 \pm 0.05 \pm 0.02 $ & $0.52 \pm 0.08 \pm 0.03$ \\
$|A_\perp|^2$ & $0.30 \pm 0.06 \pm 0.02 $ & $0.19 \pm 0.08 \pm 0.02$ \\
$\arg(A_\parallel)$(rad) & $2.39 \pm 0.24 \pm 0.04 $ & $2.10 \pm 0.28 \pm 0.04$ \\
$\arg(A_\perp)$(rad) & $2.51 \pm 0.23 \pm 0.04 $ & $2.31 \pm 0.30 \pm 0.07$ \\
\hline\hline
\end{tabular}
\end{center}
\end{table}
\begin{table}
\caption{$\Lambda$ and $\Sigma$ values obtained from the decay amplitudes
measured for $B^0$ and $\bar{B^0}$ separately.\label{bbangresults}}
\begin{center}
\begin{tabular}{c|cc}
\hline\hline
  & $B^0$ & $\bar{B^0}$\\
\hline
$|A_0|^2$ & $0.39 \pm 0.08 \pm 0.03 $ & $0.51 \pm 0.07 \pm 0.02$ \\
$|A_\perp|^2$ & $0.37 \pm 0.09 \pm 0.02 $ & $0.25 \pm 0.07 \pm 0.01$ \\
$\arg(A_\parallel)$(rad) & $2.72 ^{+ 0.46}_{-0.38} \pm 0.14 $ & $2.08 \pm 0.31 \pm 0.04$ \\
$\arg(A_\perp)$(rad) & $2.81 \pm 0.36 \pm 0.11 $ & $2.22 \pm 0.35 \pm 0.05$ \\
\hline
$A^0_T$ & $0.13 ^{+ 0.11}_{-0.14} \pm 0.04 $ & $0.28 \pm 0.08 \pm 0.01$ \\
$A^\parallel_T$ & $0.03 \pm 0.08 \pm 0.01 $ & $0.03 \pm 0.06 \pm 0.01$ \\
\hline\hline
\end{tabular}
\end{center}
\end{table}

\begin{table}
\caption{The measured decay amplitudes and triple-product correlations in the
$B^0$ and $\bar{B^0}$ samples.\label{otherresults}}
\begin{center}
\begin{tabular}{c|c}
\hline\hline
$\Lambda_{00} = 0.45 \pm 0.05 \pm 0.02$ & $\Sigma_{00} = -0.06 \pm 0.05 \pm 0.01$\\
$\Lambda_{\parallel\parallel} = 0.24 \pm 0.06 \pm 0.02$ & $\Sigma_{\parallel\parallel} = -0.01 \pm 0.06 \pm 0.01$\\
$\Lambda_{\perp\perp} = 0.31 \pm 0.06 \pm 0.01$ & $\Sigma_{\perp\perp} = -0.06 \pm 0.05 \pm 0.01$\\
$\Sigma_{\perp 0} = -0.41^{+0.16}_{-0.14} \pm 0.04$ & $\Lambda_{\perp 0} = 0.16^{+0.16}_{-0.14} \pm 0.03$ \\
$\Sigma_{\perp \parallel} = -0.06 \pm 0.10 \pm 0.01$ & $\Lambda_{\perp \parallel} = 0.01 \pm 0.10 \pm 0.02$ \\
$\Lambda_{\parallel 0} = -0.45 \pm 0.11 \pm 0.01$ & $\Sigma_{\parallel 0} = -0.11 \pm 0.11 \pm 0.02$\\
\hline\hline
\end{tabular}
\end{center}
\end{table}

The decay angles of a $B$-meson decaying to two vector mesons $\phi$ and $K^*$
are defined in the transversity basis\cite{polarization}.  The $x-y$ plane is
defined to be the decay plane of $K^*$ and the $x$ axis is in the direction
of the $\phi$ meson.  The $y$ axis is perpendicular to the $x$ axis in the decay
plane and is on the same side as the kaon from the $K^*$ decay.
The $z$ axis is perpendicular to the $x-y$ plane according to the
right-hand rule, $\theta_{\rm tr}(\phi_{\rm tr})$ is the polar(azimuthal)
angle with respect to the $z$-axis of the $K^+$ from $\phi$ decay
in the $\phi$ rest frame, and $\theta_{K^*}$ is defined earlier.

The distribution of the angles, $\theta_{K^*}$, $\theta_{\rm tr}$, and
$\phi_{\rm tr}$ is given by\cite{phikstarangle}
\bea
\frac{d^3R_{\phi K^*}(\phi_{\rm tr}, \cos\theta_{\rm tr},\cos\theta_{K^*})}
     {d\phi_{\rm tr} d\cos\theta_{\rm tr} d\cos\theta_{K^*}}  = 
   \frac{9}{32\pi}[ |A_\perp|^2 2\cos^2\theta_{\rm tr}\sin^2\theta_{K^*}
  +  |A_\parallel|^2 2\sin^2\theta_{\rm tr}\sin^2\phi_{\rm tr}\sin^2\theta_{K^*} \nonumber\\
+  |A_0|^2 4\sin^2\theta_{\rm tr}\cos^2\phi_{\rm tr}\cos^2\theta_{K^*} 
 +  \sqrt{2}{\rm Re}(A_\parallel^*A_0) \sin^2\theta_{\rm tr}\sin 2\phi_{\rm tr}\sin 2\theta_{K^*} \nonumber \\
 -  \eta \sqrt{2}{\rm Im}(A_0^*A_\perp) \sin 2\theta_{\rm tr}\cos\phi_{\rm tr}\sin 2\theta_{K^*} 
 -  2 \eta {\rm Im}(A_\parallel^*A_\perp) \sin 2\theta_{\rm tr}\sin\phi_{\rm tr}\sin^2\theta_{K^*}] \nonumber
\eea
where $A_0$, $A_\parallel$, and $A_\perp$ are the complex amplitudes of
the three helicity states in the transversity basis with the normalization
condition $|A_0|^2 + |A_\parallel|^2 + |A_\perp|^2 = 1$, and $\eta = =1(-1)$
corresponds to $B(\bar{B})$ mesons and is determined from the charge of the
kaon or pion in the $K^*$ decay.  The longitudinal polarization component
is denoted by $A_0$ and $A_\perp (A_\parallel)$ is the transverse polarization
along the $z$-axis ($y$-axis).  The value of $|A_\perp|$ $(|A_0|^2 + |A_\parallel|^2)$ is the $CP$-odd ($CP$-even) fraction in the decay
$B \to \phi K^{*0}$~\cite{phikstarangle}.
The presence of final state interactions (FSI) results in phases that differ from
either $0$ or $\pm \pi$.

The complex amplitudes are determined by performing an unbinned maximum
likelihood fit to the $B \to \phi K^*$ candidates in the signal region.  The
combined likelihood is given by
\be
\mathcal{L} = \prod^{N_{ev}}_i \epsilon( \phi_{\rm tr}, \cos\theta_{\rm tr},
\cos\theta_{ K^*}) \sum^{}_{j}f_jR_j( \phi_{\rm tr}, \cos\theta_{\rm tr},
\cos\theta_{ K^*}),
\ee
where $j$ denotes the contributions from $\phi K^*$, $q\bar{q}$,
$K^+K^-K^*$ and $f_0 K^*$; $R_j$ is the angular distribution function (ADF).
The ADF $R_{q\bar{q}}$ is determined from side-band data, and
$R_{K^+K^-K^*}$ from events with $1.04 {\rm GeV}/c^2 < M_{K^+K^-} < 
1.075 {\rm GeV}/c^2$; $R_{f_0 K^*}$ is obtained from $ B\to f_0 K^*$ MC.
The detection efficiency ($\epsilon$) is determined using MC simulations
assuming a phase space decay.  The fractions $f_j$ are parameterized
as a function of $\Delta E$ $M_{bc}$ and $M_{K^+K^-}$.  The value of $\arg(A_0)$
is set to zero and $|A_\parallel|$ is calculated from the normalization
condition.  The four parameters ($|A_0|$, $|A_\perp|$, $\arg(A_\parallel)$,
and $\arg(A_\perp)$) are determined from the fit.  There is a two-fold
ambiguity in the solutions for the phases; the chosen set of solutions
is the one suggested in the reference~\cite{unfold}.
Figure~\ref{fig:angdist} shows the angular distributions with the
projections of the fit superimposed.  The obtained amplitudes are
summarized in Table~\ref{angresults}.

The systematic uncertainties on the amplitudes are dominated by the
efficiency modeling (4-5\%), continuum background (3-4\%), slow pion
efficiency (2-3\%), and $K^+K^-K^*$ ADF (1-2\%).  The remaining possible
systematic errors such as the angular resolution, signal yields, background
from higher $K^*$ states, and width of the $f_0$ are estimated to be
less than 1\%.

The triple-product for a $B$ meson decay to two vector mesons takes
the form $\vec{q}\cdot(\vec{\epsilon_1} \times \vec{\epsilon_2})$, where
$\vec{q}$ is the momentum of one of the vector mesons, and $\vec{\epsilon_1}$
and $\vec{\epsilon_2}$ are the polarizations of the two vector mesons.
The following two $T$-odd~\cite{london,notation} quantities
\be
A^0_T = {\rm Im}(A_\perp A^*_0), A^\parallel_T = {\rm Im}(A_\perp A^*_\parallel),
\ee
provide information on the asymmetry of the triple products.  The SM
predicts very small values for $A^0_T$ and $A^\parallel_T$.  The comparison
of these triple product asymmetries ($A^0_T$ and $A^\parallel_T$) with
the corresponding quantities for the $CP$-conjugated decays
($\bar{A}^0_T$ and $\bar{A}^\parallel_T$) provides an observable sensitive
to $T$-violation.

Additional variables that can be measured through angular analysis are suggested\cite{sinha} and are given by
\bea
\Lambda_{\perp i} = - {\rm Im} (A_\perp A^*_i - \bar{A}_\perp \bar{A}^*_i), & \Lambda_{\parallel 0} = {\rm Re} (A_\parallel A^*_0 + \bar{A}_\parallel \bar{A}^*_0), \nonumber\\
\Sigma_{\perp i} = - {\rm Im} (A_\perp A^*_i + \bar{A}_\perp \bar{A}^*_i), & \Sigma_{\parallel 0} = {\rm Re} (A_\parallel A^*_0 - \bar{A}_\parallel \bar{A}^*_0), \nonumber\\
\Lambda_{\lambda\lambda} = \frac{1}{2}(|A_\lambda|^2+ |\bar{A}_\lambda|^2), &  \Sigma_{\lambda\lambda} = \frac{1}{2}(|A_\lambda|^2- |\bar{A}_\lambda|^2),  \nonumber\\
\eea
where the subscript $\lambda$ is one of 0, $\parallel$, or $\perp$ and $i$ is
one of 0 or $\parallel$.  The variables $\Lambda_{\perp 0}$ and $\Lambda_{\perp
  \parallel}$ are sensitive to $T$-violating new physics.  The following equations
should hold in the absence of NP:
\be
\Sigma_{\lambda\lambda} = 0, \Sigma_{\parallel 0} = 0, \Lambda_{\perp i} = 0.\nonumber
\ee

By separating $B^0$ and $\bar{B^0}$ samples and rearranging fitting parameters
in the unbinned maximum likelihood fit, we obtain the decay amplitudes for the
$B^0$ and $\bar{B^0}$, the triple-product correlations, and the other
NP-sensitive observables, which are given in Table~\ref{bbangresults}
and~\ref{otherresults}.

In summary improved measurements of the decay amplitudes $B\to \phi K^*$ are
presented, based on fits to angular distributions in the transversity basis.
The results are consistent with our previous measurements~\cite{chen} but
with improved precision.  The measured value of $|A_\perp|^2$ shows
that $CP$-odd ($|A_\perp|^2$) and $CP$-even($|A_0|^2 + |A_\parallel|^2$)
components are present in $\phi K^*$ decays in a ratio of 1:2.
Both phases of $A_\perp$ and  $A_\parallel$ differ from zero or $-\pi$
rad by 4:3 standard deviations ($\sigma$) which provides evidence for
the presence of final state interactions.  The measured direct $CP$ asymmetries
in these modes are consistent with zero.  These correspond to 90\% confidence
level limits of $-0.14 < A_{CP}(\phi K^{*0}(K^+\pi^-)) < 0.17$,
and $-0.25 < A_{CP}(\phi K^{*+}) < 0.22$.  Difference between triple
products asymmetries ($\Lambda_{\perp 0, \parallel}$) which are sensitive
to $T$-violation are consistent with zero.
The equations $\Sigma_{\lambda\lambda} = 0$, $\Sigma_{\parallel 0} = 0$,
and $\Lambda_{\perp i} = 0$ should be hold in the absence of NP.  Our
data does not show any significant violation of these relations.
Measurements of the $T$-violation sensitive differences between triple product
asymmetries, $A^0_T - \bar{A}^0_T$ and $A^\parallel_T - \bar{A}^\parallel_T$,
indicate no significant deviations from zero.
Our data indicates no significant
deviations from the expectations: $\Sigma_{\lambda\lambda} = 0$,
$\Sigma_{\parallel 0} = 0$, and $\Lambda_{\perp i} = 0$, indicating
no evidence for new physics.

\section{$B \to \rho K^*$ Analysis}\label{btorhok}
We select $B^+ \to \rho^+ K^{*0}$ candidate events by combining three
charged tracks (two oppositely charged pions and on kaon) and
one neutral pion.  Each charged track is
required to have a transverse momentum $p_T > 0.1 {\rm GeV}/c$ and 
be originated from the interaction point (IP).
Candidates $\pi^0$ mesons are reconstructed from pairs of photons
that have an invariant mass in the range $0.1178 {\rm GeV}/c^2 < M_{\pi^0}
< 0.1502 {\rm GeV}/c^2$.  The $\pi^0$ candidates are kinetically
constrained to the nominal $\pi^0$ mass.  In order to reduce the combinatorial
background, we only accept $\pi^0$ candidates with momenta $p_{\pi^0} > 0.40
{\rm GeV}/c$ in the cms.
Candidate $\rho^+$ mesons are reconstructed via their $\rho^+ \to \pi^+\pi^0$
decay, and the $\pi^+\pi^0$ pairs are required to have an invariant mass
in the region $0.62 {\rm GeV}/c^2 < M_{\pi^+\pi^0}< 0.92 {\rm GeV}/c^2$.
Candidate $K^{*0}$ mesons are selected from the $K^{*0} \to K^+\pi^-$
decay with an invariant mass $0.83 {\rm GeV}/c^2 < M_{K^+\pi^-}< 0.97
{\rm GeV}/c^2$.  
To isolate the signal we accept events in the region $M_{bc}>5.2{\rm GeV}/c^2$
and $-0.3 {\rm GeV} < \Delta E < 0.3 {\rm GeV}$, and define a signal region
in $M_{bc}$ and $\Delta E$ as $M_{bc}>5.27{\rm GeV}/c^2$ and
$-0.10 {\rm GeV} < \Delta E < 0.06 {\rm GeV}$ respectively.
The continuum process is the main source of background to be suppressed.
In addition to the continuum process reduction at $B \to
\phi K^*$ analysis, the displacement along the beam direction between the signal
$B$ vertex and that of the other $B$, $\delta z$, also provides separation.
For $B$ events, the average value of $\delta z$ is approximately $200 \mu m$
while continuum events have a common vertex.  This suppression
removes 99.3\% of the continuum background while retaining 41 \% of the
$B^+ \to \rho^+K^{*0}$ events.  The MC-determined efficiency with all selection
criteria imposed is 2.7\% for longitudinal polarization($A_0$) and 4.0\%
for transverse polarization ($A_\pm$).

To investigate backgrounds from $b \to c $ decays, we use a sample of $B\bar{B}$
MC events corresponding to an integrated luminosity of $412 {\rm fb}^{-1}$.
We find a contribution from $B^+ \to \bar{D^0}(K^+\pi^-\pi^0)\pi^+$ decays
in the $\rho$ or $K^*$ sideband region and require$|M_{K\pi\pi^0}-M_{D^0}|
> 0.050 {\rm GeV}/c^2$ to veto these events.  
Among the charmless $B$ decays,
potential background arise from $B^+ \to a_1^0 K^+$, $B^+ \to \rho^+K^{*0}(1430)$,
non-resonant $B^+ \to \rho^+K^+\pi^-$ and $B^+ \to K^{*0}\pi^+\pi^0$.
We separate signal from these backgrounds by fitting the $\rho$ and $K^*$
invariant mass distributions.

\begin{figure}
  \begin{minipage}{0.45\hsize}
  \begin{center}
  \psfig{figure=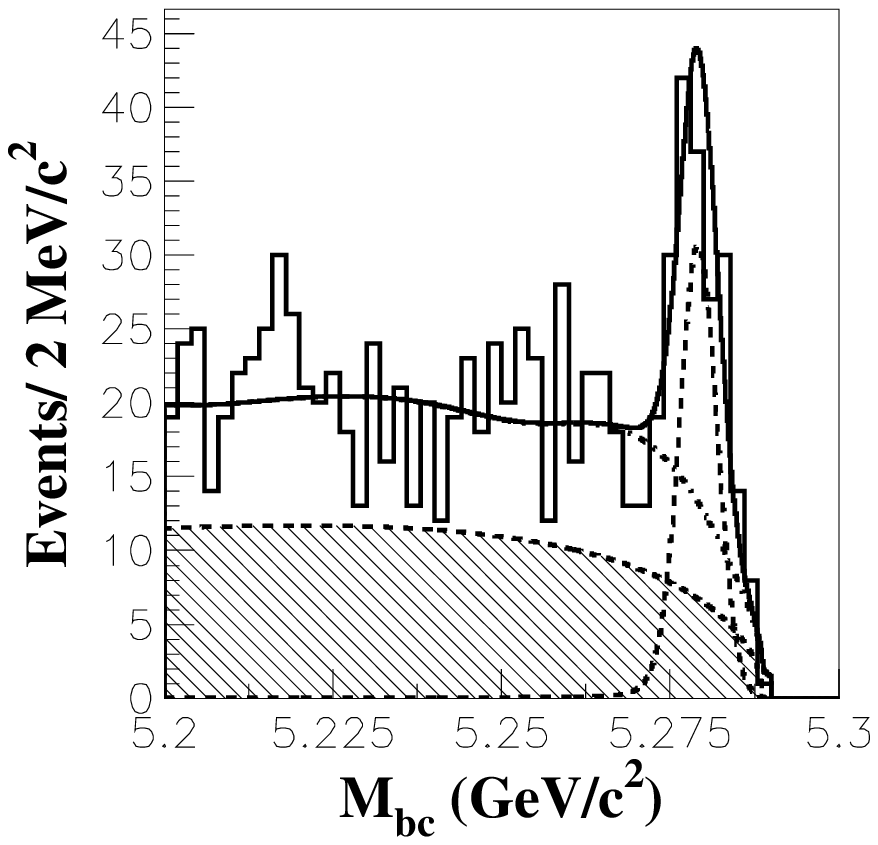,height=1.3in}
  \psfig{figure=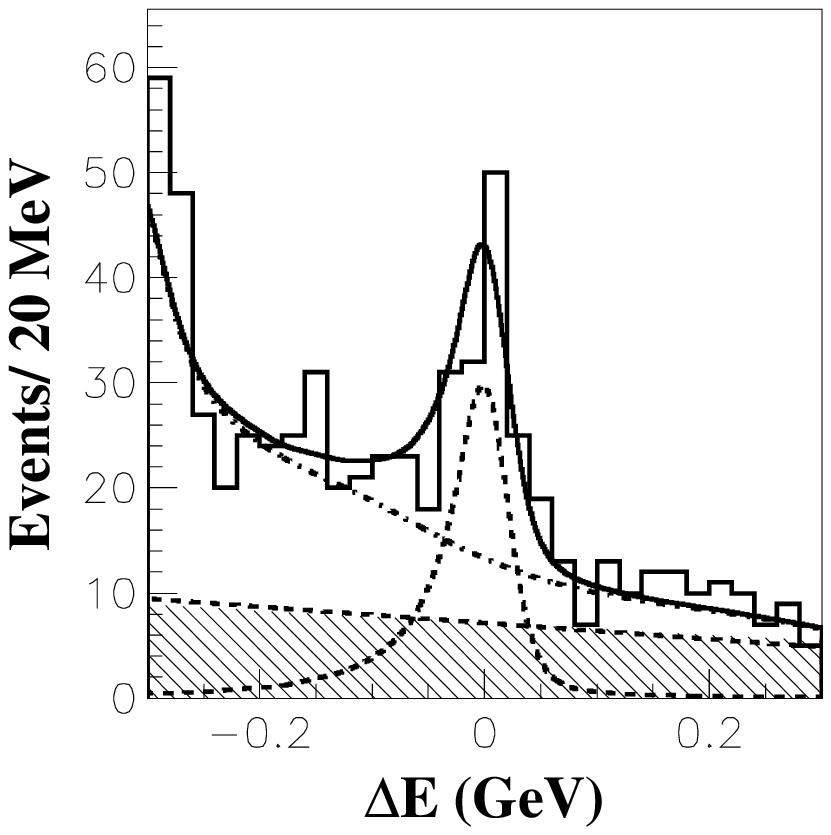,height=1.3in}
\caption{Projections of $M_{bc}$ for events in the $\Delta E$ signal
  region (left), and projection of $\Delta E$ in the $M_{bc}$ signal
  region (right).  The solid curves show the results of the fit.  The
  hatched histograms represent the continuum background.  The sum of
  the $b\to c$ and continuum background component is shown as
  dot-dashed lines.
\label{fig:rhokstar}}
  \end{center}
  \end{minipage}
  \begin{minipage}{0.45\hsize}
  \begin{center}
  \psfig{figure=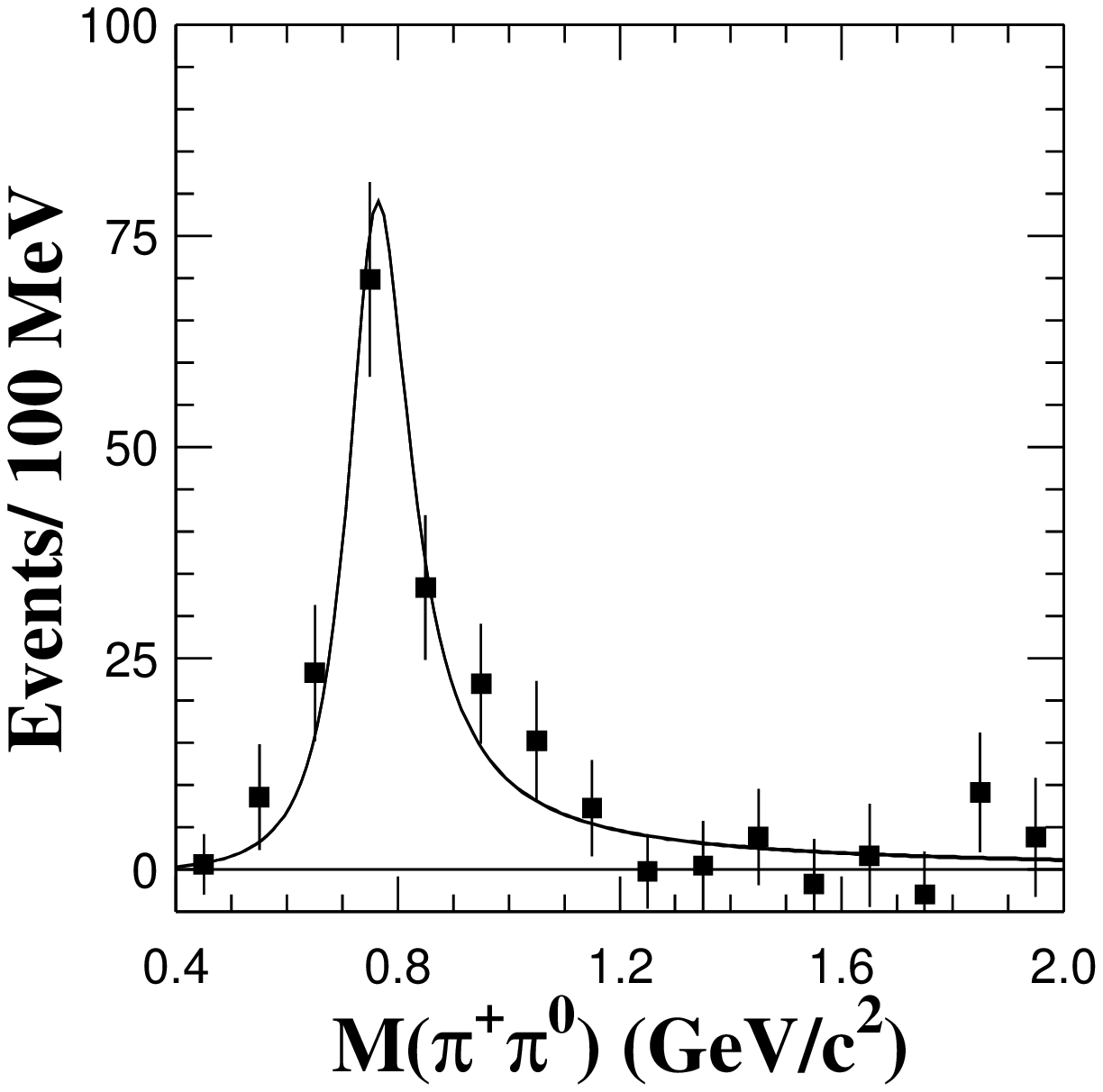,height=1.3in}
  \psfig{figure=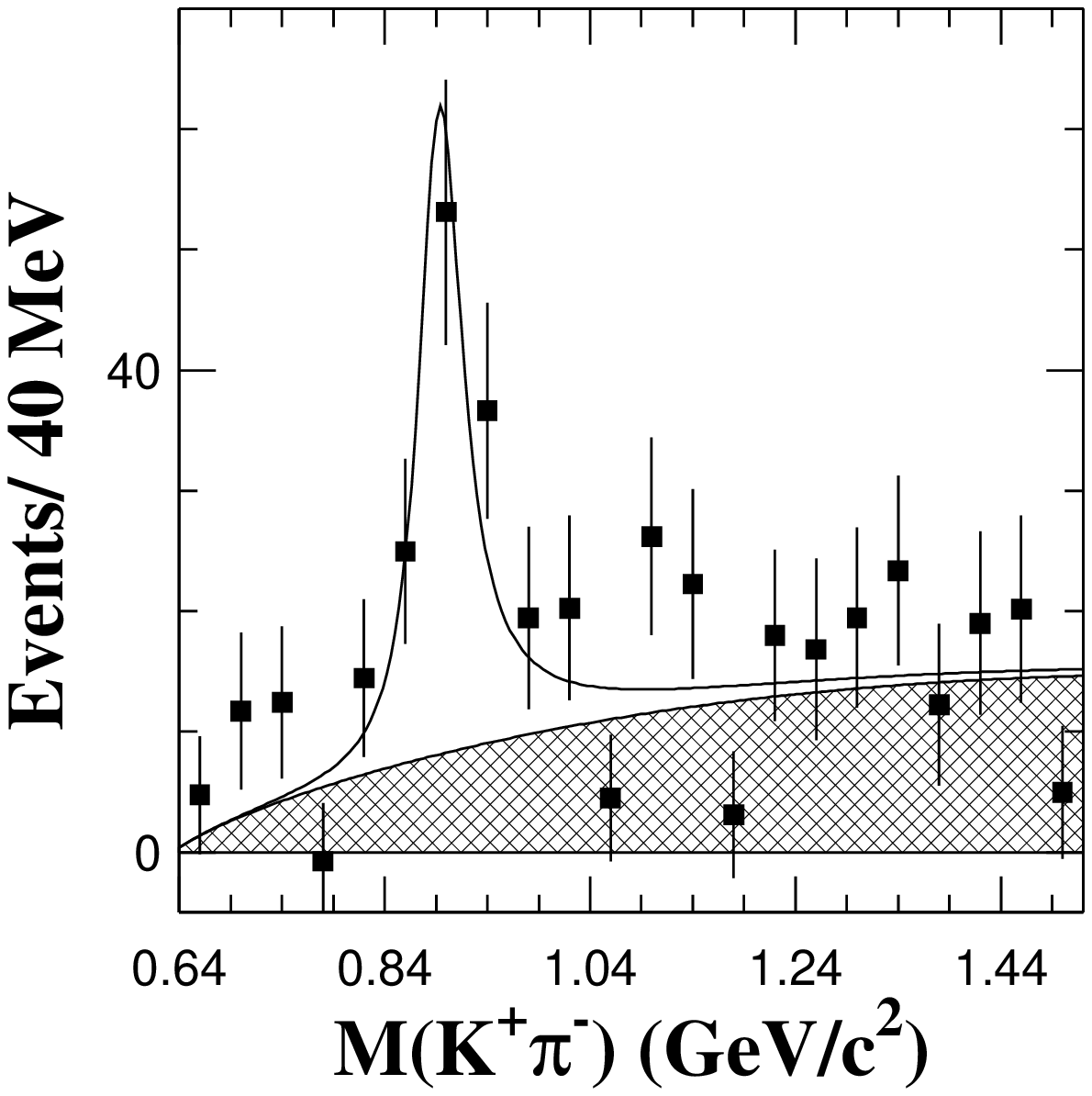,height=1.3in}
\caption{Signal yields obtained from the $M_{bc}-\Delta E$ distribution
  in bins of $M_{\pi^+\pi^0}$ (left) for events in the $K^{*0}$ region
  and in bins of $M_{K\pi}$ (right) for events in the $\rho$ region.
  Solid curves show the results of the fit.  Hatched histograms are
  for the non-resonant component.
  \label{fig:rhokstar2}}
  \end{center}
  \end{minipage}
\end{figure}

We extract the signal yield by applying an extended unbinned maximum-likelihood
fit to the two-dimensional $M_{bc}-\Delta E$ distribution.  The fit includes
components for signal plus backgrounds from continuum events and $b\to c$
decays.  The PDFs for signal and $b\to c$ decay are modeled by smoothed
two-dimensional histograms obtained from large MC samples.  The signal PDF
is adjusted to account for small differences observed between data and MC
for a high-statistics mode containing $\pi^0$ mesons, $B^+ \to \bar{D^0}
(K^+\pi^-\pi^0)\pi^+$.  The continuum PDF is described by a product of a
threshold (ARGUS) function for $M_{bc}$ and a first-order polynomial for
$\Delta E$, with shape parameters allowed to vary.  All normalizations are
allowed to float.  Figure~\ref{fig:rhokstar} shows the final event sample
and the fit results.   The five-parameter (three normalizations plus
two shape parameters for continuum) fit yields $134.8 \pm 16.9$
$B^+ \to K^+\pi^-\pi^+\pi^0$ events.

We further distinguish the $\rho^+K^{*0}$ signal from non-resonant decays
such as $B^+ \to \rho^+K^+\pi^-$ or $B^+ \to K^{*0}\pi^+\pi^0$ by fitting
the $M_{\pi^+\pi^0}$ and $M_{K\pi}$ invariant mass distributions.  The signal
yields obtained from the $M_{bc}-\Delta E$ fit for different
$M_{\pi^+\pi^0}$ and $M_{K\pi}$ bins are plotted in Figure~\ref{fig:rhokstar2},
where the $M_{\pi^+\pi^0}$ distribution is for events in the $K^{*0}$ region
($0.83 {\rm GeV}/c^2 < M_{K\pi} < 0.97 {\rm GeV}/c^2$) and the
the $M_{K\pi}$ distribution is for events in the $\rho$ region
($0.62 {\rm GeV}/c^2 < M_{\pi^+\pi^0} < 0.92 {\rm GeV}/c^2$).
We perform separate $\chi^2$ fits to the $M_{\pi\pi}$ or $M_{K\pi}$
distributions.  Each fit includes components for signal and non-resonant
background.  the signal $\rho$ and $K^*$ PDFs are modeled by relativistic
$P$-wave Breit-Wigner functions with means and widths fixed at their known
values~\cite{f02}; the PDFs are convolved with a Gaussian of $\sigma = 5.3
{\rm MeV}$, which is obtained by fitting the $D^0(K^-\pi^+)$ invariant
mass, to account for the detector resolution.  The non-resonant component
is represented by a threshold function with parameters determined from MC
events where the final states are distributed uniformly over phase space.
The $M_{\pi\pi}$ mass fit gives $125.4 \pm 15.8 \rho$ and $-0.3 \pm 3.0$
non-resonant events in the $K^*$ mass region.  The statistical significance
of the signal, defined as $\sqrt{\chi^2_0 - \chi^2_{min}}$, where
$\chi^2_{min}$ is the $\chi^2$ value at the best-fit signal yield and
$\chi^2_0$ is the value with the $K^{*0}$ signal yield set to zero,
is $5.3 \sigma$($5.2\sigma$ with the inclusion of systematics).
The contribution from non-resonant $\rho^+K^+\pi^-$ is significant and
is taken into account in both the branching fraction and polarization
determinations, while we neglect the non-resonant $K^{*0}\pi^+\pi^0$
contribution.

\begin{figure}
  \begin{minipage}{0.5\hsize}
  \begin{center}
  \psfig{figure=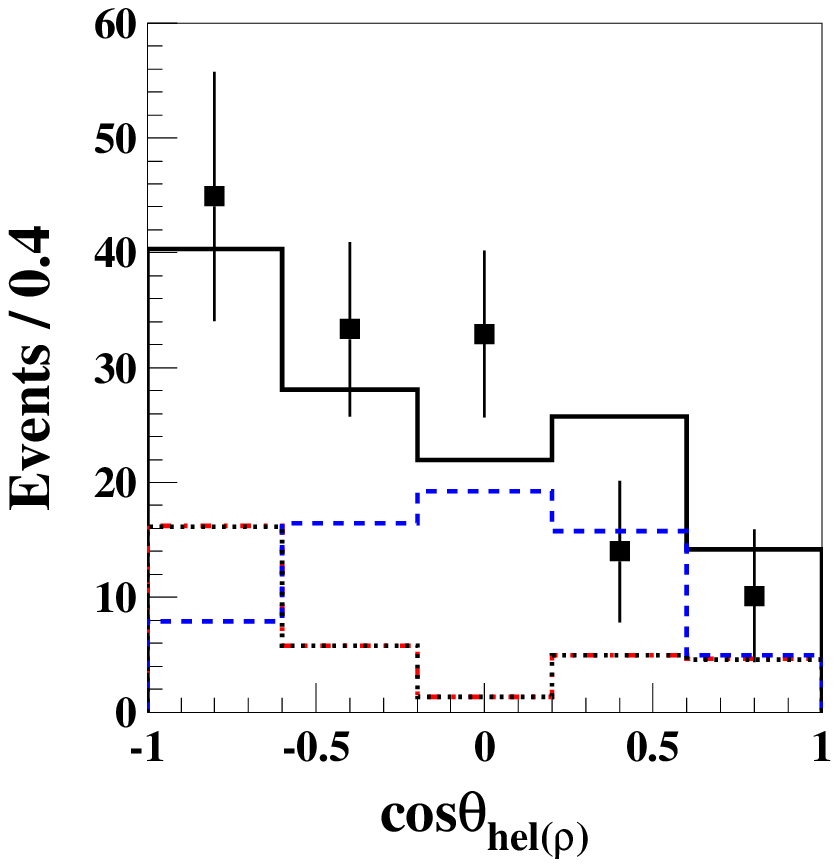,height=1.5in}
  \end{center}
  \end{minipage}
  \begin{minipage}{0.5\hsize}
  \begin{center}
  \psfig{figure=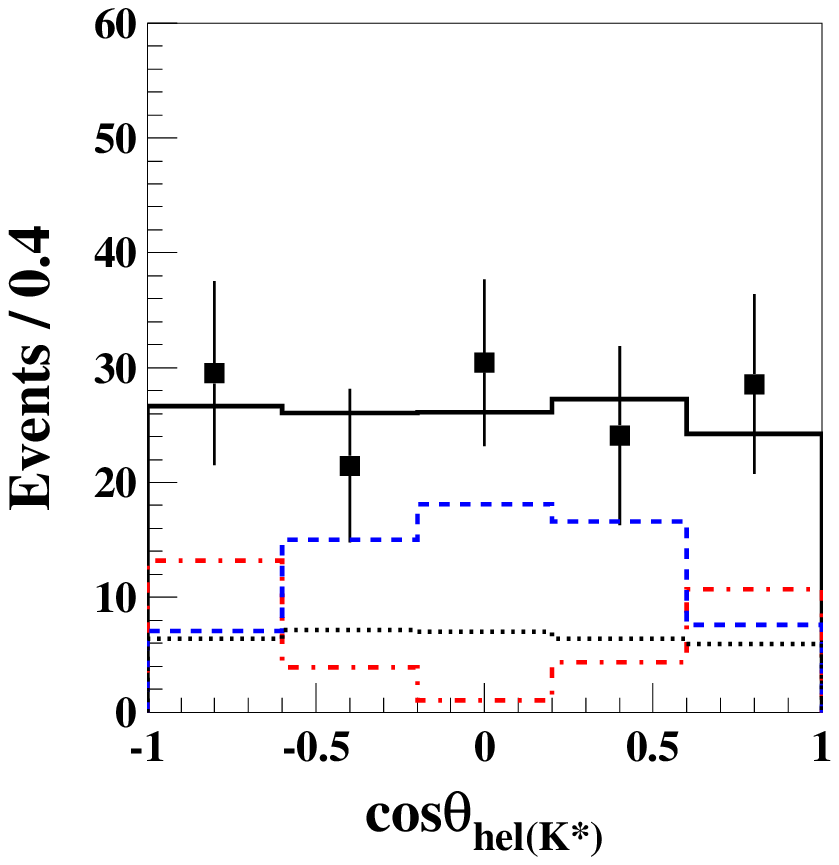,height=1.5in}
  \end{center}
  \end{minipage}
\caption{Fit to background-subtracted helicity distributions.  The solid
  histograms show the results.  The dot-dashed (dashed) histograms are the
  $A_0(A_\pm)$ component of the fit; the dotted histograms are for non-resonant
  $\rho K \pi$.  The low event near $\cos \theta_\rho = 1 $ is due to the
  $p_{\pi^0} > 0.4 {\rm GeV}/c$ requirement.  
  \label{fig:rhokstarhel}}
\end{figure}

We use the $\rho^+\to \pi^+\pi^0$ and $K^{*0}\to K^+\pi^-$ helicity-angle
($\theta_\rho, \theta_{K^*}$) distributions to determine the relative
strengths of $|A_0|^2$ and $|A_\pm|^2$.  Here $\theta_\rho(\theta_{K^*})$ is
the angle between an axis anti-parallel to the $B$ flight direction and
the $\pi^+(K^+)$ flight direction in the $\rho(K^*)$ rest frame.  For
the longitudinal polarization case, the distribution is proportional to
$\cos^2\theta_\rho \cos^2\theta_{K^*}$, and for the transverse polarization case,
it is proportional to $\sin^2\theta_\rho \sin^2\theta_{K^*}$~\cite{phikstarangle}.
Figure~\ref{fig:rhokstarhel} shows the signal yields obtained from
$M_{bc}-\Delta E$ fits in bins of the cosine of the helicity angle
for $\rho$ and $K^*$.  We perform a binned simultaneous $\chi^2$ fit to
the $\rho$ and $K^*$ helicity-angle distributions.  The fit includes components
for signal and non-resonant $\rho K \pi$.  PDFs for signal $A_0$ and
$A_\pm$ helicity states are determined from the MC simulation.  The
helicity-angle distribution for data in the high $M_{K\pi}$ sideband region
$1.1 {\rm GeV}/c^2 < M_{K\pi} < 1.5 {\rm GeV}/c^2$, where $\rho K\pi$ events
dominate, is consistent with a $\cos^2\theta$-like $\cos\theta_\rho$ and
a flat $\cos\theta_{K^*}$ distribution.  Thus, we assume an $S$-wave $K\pi$
system and model the non-resonant  $B\to \rho K \pi$ PDF based on the MC
simulation.  The fraction of the non-resonant component is fixed at the values
obtained from the $K^*$ mass fit.  The two parameter (normalizations for $A_0$
and $A_\pm$) fit result deviates from 100\% longitudinal polarization with
a significance of $4.9\sigma$ ($4.3\sigma$ including systematic uncertainties).
The significance is defined as $\sqrt{\chi^2_0 - \chi^2_{\rm min}}$, where
$\chi^2_{\rm min}$ is the $\chi^2$ value at the best-fit and $\chi^2_0$ is
the value with the longitudinal polarization fraction set to 100\%.

The largest uncertainties in the polarization measurement are due to
uncertainties in the non-resonant $\rho K\pi$ PDF, potential scalar-pseudoscalar
($S-P$) interference, and the non-resonant fraction.  We assign $^{+10.3}_{-0}\%$
systematic error for the non-resonant PDF.  This uncertainty is estimated
by adding a $1/3$ flat component to the $\rho$ helicity PDF for non-resonant
$\rho K \pi$ in the helicity fit.  Interference of the longitudinal
amplitude $A_0$ with the $S$-wave ($K\pi$) system introduces a term with
a $2e^{i\Delta \phi}|A_{\rho K \pi}| \cos\theta_{K^*}$ dependence,
where $\Delta \phi$ is the phase difference and $|A_{\rho K\pi}|$ is amplitude
of the $B\to \rho K\pi$ decay.  The $S-P$ wave interference disappears in
the $\cos\theta_\rho$ distribution, which is integrated over $\cos\theta_{K^*}$;
however it remains in the $\cos\theta_{K^*}$ distribution.
We include an additional linear function for the interference term in the
$\cos\theta_{K^*}$ helicity and redo the $\chi^2$ fit.  The resulting small
change in $f_L$, 0.5\%, is assigned as the systematic uncertainty for
the $S-P$ interference.  A $^{+4.0}_{-4.1}\%$ systematic error is assigned
for the uncertainty in the fraction of non-resonant $\rho K\pi$, obtained
by varying the non-resonant fraction by $\pm 1\sigma$.  Adding the various
systematic error contributions in quadrature, we obtain the longitudinal
polarization fraction in $B^+ \to \rho^+ K^{*0} $ decays,
\be
f_L(B^+ \to \rho^+ K^{*0}) = 0.43 \pm 0.11({\rm stat.}) ^{+0.05}_{-0.02}({\rm syst.}).\nonumber
\ee

To calculate the $B^+ \to \rho^+ K^{*0}$ branching fraction, we use
the invariant mass fit result and MC-determined efficiencies weighted by the
measured polarization components.  We consider systematic errors in the
branching fraction that are caused by uncertainties in the efficiencies
of track finding, particle identification, $\pi^0$ reconstruction, continuum
suppression, fitting, polarization fraction.  We assign an error of 1.1\%
per track for the uncertainty in the track efficiency.  This uncertainty
is obtained from a study of partially reconstructed $D^*$ decays.
We also assign an uncertainty of 0.7\% per track on the particle identification
efficiency, based on a study of kinematically selected
$D^{*+} \to D^0(K^-\pi^+)\pi^+$ decay.  A 4.0\% systematic error for the
uncertainty in the $\pi^0$ detection efficiency is determined from
data-MC comparisons of $\eta \to \pi^0\pi^0\pi^0$ with $\eta \to \pi^+\pi^-\pi^0$
and $\eta \to \gamma\gamma$.  A 4.5\% systematic error for continuum
suppression is estimated from studying the process
$B^+\to\bar{D^0}(K^+\pi^-\pi^0)\pi^+$.  A -4.2\%/+1.7\% error due to the
uncertainty in the fraction of longitudinal polarization is obtained
by varying $f_L$ by its errors.  The uncertainty in non-resonant
$K^*\pi\pi$ background gives a contribution of -2.2\%/+0\% in addition to
-3.0\%/+2.3\% error from uncertainties in the background from other rare
$B$ decays.  A 1.1\% error for the uncertainty in the number of $B\bar{B}$
events in the data sample.  A 7.1\% error for possible bias in the $\chi^2$
fit~\cite{chi2}, obtained from a MC study is also included.
The quadratic sum of all of these errors is taken as the total systematic
error.  We obtain the branching fraction
\be
\mathcal{B}(B^+\to \rho^+K^{*0}) = (8.9 \pm 1.7({\rm stat}) \pm 1.2({\rm syst})
\times 10^{-6}.\nonumber
\ee

In summary, we have observed the $B^+\to \rho^+K^{*0}$ decay with a statistical
significance of $5.3\sigma$.  We measure the branching fraction to be
$(8.9 \pm 1.7({\rm stat}) \pm 1.2({\rm syst})\times 10^{-6}$.  We also
perform a helicity analysis and find a substantial transversely polarized
fraction with a statistical significance of $4.9\sigma$.  The longitudinal
polarization fraction $f_L$ measured is similar to the surprisingly low value
found in $b \to s\bar{s}s$ decays $B\to \phi K^*$.


\begin{thebibliography}{99}
\bibitem{notation} Definisions are $A^0_T = A^{(1)}_T$, and $A^\parallel_T = A^{(2)}_T$, but $\bar{A^0_T} = -\bar{A}^{(1)}_T$, and $\bar{A}^\parallel_T = -\bar{A}^{(2)}_T$~\cite{london}.
\end{thebibliography}
\end{document}